\documentstyle[aps]{revtex}
\newcommand{\ds}{\displaystyle}
\newcommand{\dsf}{\ds\frac}

\newcommand{\beq}{\begin{equation}}
\newcommand{\eeq}{\end{equation}}

\large

\begin{document}
\large

\begin{center}
\Large\bf

Nonlinear Thermomagnetic Waves in the Resistive State of Superconductors
\vskip 0.1cm
{\normalsize\bf N.А.\,Taylanov}\\
\vskip 0.1cm
{\large\em Research Institute of Applied Physics,\\
 National University of Uzbekistan,\\
Vuzgorodok, Tashkent, 700174 Uzbekistan,\\
e-mail:taylanov@iaph.tkt.uz}
\end{center}
\begin{center}
\bf Abstract
\end{center}

\begin{center}
\mbox{\parbox{14cm}{\small
        The evolution of thermomagnetic perturbations in the resistive state
of superconductors is considered. A qualitative pattern of the formation and
further development of nonlinear stationary structures that describe the final
stage of thermal and electromagnetic perturbations in a superconductor is
investigated. The wave propagation velocity and the wave front width in a
superconductor are estimated.
}}
\end{center}
\vskip 0.5cm

        Energy dissipation during the motion of vortices leads to ohmic
heating of a superconductor. As a result, a certain region of the
superconductor is heated to a temperature  $T>T_{c}$, where $T_c$ is the
critical temperature. An increase in the temperature in a local region
of the sample brings about a decrease in the critical current $j_{c}$ and
the emergence of vortex electric field $E$ at the same region.
        The effect of superconductivity quenching due to thermal heating
of the vortex lattice has been experimentally studied for a long time. Early
experiments [I] revealed that the vortex electric field E is induced under
ohmic heating in a superconductor through which a direct current with a
density $j_c$ flows. According to the balance between dissipative and
nonlinear effects, the transition to the resistive state is accompanied
by the emergence of various modes of the "switching wave" type, i.e.,
the regime of a wave motion that switches a sample from the superconducting
to the normal state. Examples of these modes can be provided by thermal
waves, namely, a steady-state propagation of a normal zone [2] or nonlinear
thermomagnetic waves [3] in superconductors.

        In this work, we studied the qualitative pattern of formation and the
profile of nonlinear dissipative structures, i.e., stationary traveling
waves which describe the final stage of the evolution of thermal and
electromagnetic perturbations in the resistive state of superconductors.

        The evolution of thermal and electromagnetic perturbations in a
superconductor is described by the nonlinear one-dimensional heat
conduction equation

\beq
\nu\dsf{dT}{dt}=\kappa\dsf{d^2T}{dx^2}+\vec j\vec E
\eeq
the Maxwell equation
\beq
\dsf{d^2\vec E}{dx^2}=\dsf{4\pi}{c^2}\dsf{d\vec j}{dt}\,,
\eeq
and their related equation of the critical state
\beq
\vec j=\vec j_{с}(T,\vec H)+\vec j_{r}(\vec E)\,.
\eeq
where $\nu$ is the heat capacity and $\kappa$ is the thermal conductivity
coefficients; $\vec j_{c}$ and $\vec j_{r}$ are the densites of the critical
and resistive currents, respectively.

The model under consideration is essentially nonlinear, because the
right-hand side of Eq. (1) contains a term describing the Joule heat
generation in the region of the resistive phase. An exact solution to
the essentially nonlinear parabolic partial differential equations
(1)-(3) does not exist.

Note that the evolution of perturbations of the temperature $T(x,t)$
and the fields $E(x,t)$ and $H(x,t)$ is determined by the equation of
critical state (3).
Because of considerable analytical difficulties,
we will restrict ourselves to considering the Bean model [4] and
assume that the critical current density is independent of the external
magnetic field; i.e., $\dsf{dj_{c}}{dH}=0$. The dependence $j_c(T)$ is
described by the relationship $j_{c}(T)=j_{0}-a(T-T_0)$; where $T_ 0$
is the initial temperature of the superconductor and the quantity
$a=\left|\dsf{dj_c}{dT}\right|_{T=T_0}$
describes the thermally activated weakening of the Abrikosov vortex
pinning by lattice defects. The dependence $j_r(E)$ in the region of the
electric field $E\ge E_f$ ($E_f$ is the boundary of the linear section in the
current-voltage characteristic of the superconductor) can be approximated
by a piecewise-linear function $j_r(E)\sim \sigma_{f} E$, where $\sigma_f$ is
the effective conductivity. The dependence $j_r(E)$ is essentially nonlinear
in the flux creep region $E\le E_f$ [5]. Here, we will consider a perturbation
with a sufficiently high amplitude and use the linear dependence
$j_{r}(E)$.

We seek a solution to the initial system of equations as a function
of the new self-simulated variable $\xi(x,t)$, that is
\beq
\begin{array}{ll}
T=\Theta[\xi(x,t)]\,,
E=E[\xi(x,t)]\,,
H=H[\xi(x,t)]\,,
\end{array}
\eeq
which $\xi=x-vt$ describes a traveling wave that moves with a constant velocity
$v$ along the $x$ axis [3].

After substitution of relationships (4) into the initial system et of equations
and performing a trivial differentation, we obtain the following system of equations
for the variable $\xi(x,t)$

\beq
- \nu v\dsf{dT}{d\xi}=\dsf{d}{d\xi}\left[\kappa\dsf{dT}{d\xi}\right]+jE,
\eeq
\beq
\dsf{dE}{d\xi}=-\dsf{4\pi v}{c^2}j,
\eeq
\beq
E=\dsf{v}{c}H.
\eeq

The corresponding thermal and electrodynamic boundary conditions for
Eqs. (5)-(7) have the form

\beq
\begin{array}{ll}
T(\xi\rightarrow+\infty)=T_0\,,&\quad \dsf{dT}{dx}(\xi\rightarrow-\infty)=0\,,\\
E(\xi\rightarrow+\infty)=0\,,&\quad   E(\xi\rightarrow-\infty)=E_e\,.
\end{array}
\eeq
where $E_e$ is the amplitude of the electric field.
It should be noted that inclusion of the temperature dependences of the
parameters $\kappa$ and $\nu$ substantially complicates analytical
calculations of the wave evolution dynamics that is described by the
system of Eqs. (5)- (7). In most cases, the changes in the local values of these
parameters in the sample can be considered negligible comparatively to the
characteristic scale of temperature variations. Hence, we can take
these parameters to be constant.
Indeed, the investigation revealed that the thermal
conductivity almost does not affect the character of the stationary
wave propagation. This stems from the fact that the thermal flux
$\kappa\dsf{dT}{d\xi}$ vanishes at stationary points of the system
at $\xi\rightarrow\pm \infty$. However, the
temperature dependence of the heat capacity should be taken into account.
Such a dependence is representeq as
$\nu\approx\nu_0\left(\dsf{T}{T_0}\right)$ over a wide range of
temperatures [5].

By eliminating variables $T(x,t)$ and $H(x,t)$ with the aid of relationships
(5) and (7) and employing the boundary conditions (8), we obtain a
differential equation for the $E$ wave distribution:

\beq
\frac{d^2 E}{dz^2}-2\pi\frac{\nu T_0}{E_\kappa}\frac{v^2}{c^2}\left[\left(1+
\frac{\sigma_f E}{aT_0}+\frac{c^2}{4\pi avT_0}\frac{dE}{dz}\right)^{4}-1
\right]+
\beta\tau\frac{dE}{dz}=\frac{E^2}{2E_\kappa}\,.
\eeq
Here, we introduced the following dimensionless parameters:
$$
z=\dsf{\xi}{L}\,,\quad
L=\frac{cH_e}{4\pi j_0}\,,\quad
E_\kappa=\frac{\kappa}{aL^2}\,,\quad
\beta=\frac{vt_\kappa}{L}\,,\quad
\tau=\frac{4\pi\sigma_f\kappa}{c^2\nu}\,,\quad
t_\kappa=\frac{\nu L^2}{\kappa}\,,
$$
where $L$ is the depth of the magnetic field penetration into the sample
and $t_\kappa$ is the thermal time of the problem.
        According to the qualitative theory [6], the equilibrium states are
found from the condition
\beq
2\pi\nu_0 T_0\frac{v^2}{c^2}
\left[\left(1+\frac{\sigma_f E}{aT_0}\right)^{4}-1\right]=E^2\,.
\eeq

An evident property of system ( 10) is the absence of closed curves that
are fully composed of the phase trajectories in the phase plane
$\left(E,\dsf{dE}{d\xi}\right)$.
The proof of this statement can be based on the Bendixson criterion [7] .
The number of stationary points ( one or three) and their
type are determined by the parameter
\beq
P=2\pi\nu_0T_0\frac{v^2}{c^2}\,.
\eeq

The three equilibrium points $E=0$, $E=E_1$ and $E=E_2$ correspond to the
condition $P<P_k=\dsf{1}{2}$ (Fig.1). There is only one singular point
$E_0=0$ at $P>P_k$. The parabola and the quartic curve in Eq. (10) are
tangent at $P=P_k$; i.e., this condition corresponds to the coincidence
$E_1=E_2=E^*=\dsf{6}{7}\frac{aT_0}{\sigma_{f}}$.
The direct solution of Eq. (10) yields the following waves:
\begin{eqnarray}
\begin{array}{lll}
1)& E_{1,2}=E^*[1+2,2(P_k-P)^{1/2}]\,,&\quad\mbox{при}\quad
\left(\dsf{P_k-P}{P_k}\right)<<1\,;\\
\quad\\
2)& E_1=8\pi\dsf{\sigma_f\nu_0}{a} \frac{v^2}{c^2}\,,&\\
\quad\\
  & E_2=(2\pi)^{1/2} \dsf{\sigma_f^2\nu_0^2}{a^2}\frac{c}{v}>>E_1\,,&
\quad\mbox{at}\quad P_k>>P\,.
\end{array}
\end{eqnarray}
       Analysis of the phase plane shows that the points $E_0=0$ and $E=E_2$
are stable nodes and that $E=E_1$ is a saddle. In addition to the separatrix
$E_1 E_0$, the set (10) has the separatrix $E_1 E_2$ connecting the points
$E_1$ and $E_2E_1$ (Fig.2). This means that two types of waves with amplitudes
$\Delta E=E_1$ and $\Delta E=E_2-E_1$ can exist in the superconductor.
Evidently, wave 1) has an amplitude of the order $E_k$ at $P\rightarrow P_k$;
its velocity is determined by equality (10) at
$E=E_{1}$. Equation (10) has two stationary points at $P<<P_k$:  $E_{0}=0$
is a stable node and $E_{1}=2\beta^2\tau E_{\kappa}$ is a saddle.

The separatrix that connects these two equilibrium states corresponds to a
"difference"-type solution with amplitude $E_{e}$, which is related to the wave
velocity  $v_{E}$ and the wave front width $\Delta z$ by the following
equations:

\beq
v_{E}=\frac{L}{t_{\kappa}}\left[\dsf{E_{e}}{2\tau E_{\kappa}}\right]^{1/2}\,,
\eeq
\beq
\Delta z=16\dsf{1+\tau}{\tau^{1/2}}\left[\dsf{E_{\kappa}}{E_{e}}\right]^{1/2}\,.
\eeq
 Wave 2) has a small amplitude at $P\rightarrow P_k$
\beq
\frac{\Delta E}{E_k}=4,4(P_k-P)^{1/2} <<1\,,
\eeq
and its velocity is inversely proportional to the amplitude at $P<<P_k$.
Such an exotic dependence of $v_E$ on $\Delta E =E_e$ most likely means that
the waves of this type are unstable. Note that observation of the
second-type waves becomes possible in finite-sized samples with asymmetric
boundary conditions.
In conclusion, it should be noted that the above investigations prove the
possibility of applying the results obtained to high-temperature
superconductors cooled to liquid-nitrogen temperatures (T = 77 K), providing
that the values of the physical parameters of the sample are known.

\begin{center}
{\bf REFERENCE}
\end{center}
\begin{enumerate}
\item
V.A.Al'tov, V.B.Zenkevich, M.G.Kremlev, and V.V.Sychev,
Stabilization of Superconducting Magnetic Systems (Energoatomizdat, Moscow,
1984).
\item
A.V.Gurevich and R.G.Mints, eat Autowaves in Normal Metals and
Superconductors (Nauka, Moscow, 1987).
\item
N.A.Taylanov Superconduct. Science and Technology, 14, 326 (2001).
\item
C. P. Bean, Phys. Rev. Lett. 8 (6),250 (1962).
\item
R. G. Mints and A. L. Rakhmanov, Instabilities in Superconductors
(Nauka, Moscow, 1984).
\item
V. I. Karpman, Nonlinear Waves in Dispersive Media
(Nauka, Moscow, 1973; Pergarnon, Oxford, 1975).
\item
A.A.Andronov, A.A.Vitt, S.E.Khaykin, Theory of Oscillators
(Nauka, Moscow, 1981).
\end{enumerate}

\end{document}